**Super-harmonic injection locking of nano-contact spin-torque vortex oscillators**

P. S. Keatley[1]*, S. R. Sani[2,3,4], G. Hrkac[5], S. M. Mohseni[2,6], P. Dürrenfeld[7], J. Åkerman[2,3,7], and R. J. Hicken[1]

[1] *Department of Physics and Astronomy, University of Exeter, Stocker Road, Exeter, EX4 4QL, UK.*

[2] *Materials and Nano Physics, School of ICT, KTH Royal Institute of Technology, Electrum 229, 164 60 Kista, Sweden.*

[3] *NanOsc AB, Electrum 205, 164 40 Kista, Sweden.*

[4] *Department of Physics & Astronomy, Uppsala University, Box 516, SE-751 20 Uppsala, Sweden.*

[5] *College of Engineering, Mathematics and Physical Science, University of Exeter, Exeter, EX4 4SB, UK.*

[6] *Department of Physics, Shahid Beheshti University, Tehran, Iran*

[7] *Physics Department, University of Gothenburg, Fysikgränd 3, 412 96 Gothenburg, Sweden*

# ABSTRACT

Super-harmonic injection locking of single nano-contact (NC) spin-torque vortex oscillators (STVOs) subject to a small microwave current has been explored. Frequency locking was observed up to the fourth harmonic of the STVO fundamental frequency $f_0$ in microwave magneto-electronic measurements. The large frequency tunability of the STVO with respect to $f_0$ allowed the device to be locked to multiple sub-harmonics of the microwave frequency $f_{RF}$, or to the same sub-harmonic over a wide range of $f_{RF}$ by tuning the DC current. In general, analysis of the locking range, linewidth, and amplitude showed that the locking efficiency decreased as the harmonic number increased, as expected for harmonic synchronization of a non-linear oscillator. Time-resolved scanning Kerr microscopy (TRSKM) revealed significant differences in the spatial character of the magnetization dynamics of states locked to the fundamental and harmonic frequencies, suggesting significant differences in the core trajectories within the same device. Super-harmonic injection locking of a NC-STVO may open up possibilities for devices such as nanoscale frequency dividers, while differences in the core trajectory may allow mutual synchronisation to be achieved in multi-oscillator networks by tuning the spatial character of the dynamics within shared magnetic layers.

## I. INTRODUCTION

Since the prediction[1,2] and realisation[3,4] of spin transfer torque (STT) there have been significant advances in order to utilize the effect in nanoscale non-linear microwave oscillators.[5,6] In such a device STT acts to counteract and balance the magnetic damping of a ferromagnetic resonance[7] allowing for autonomous (auto-) oscillations of the magnetization to be realized.[8,9] These so-called spin-torque oscillators (STOs) are attractive for nanoscale STO-based devices such as broadband microwave generators owing to their high frequency tunability (100 MHz to 100 GHz).[10,11] The microwave emission of STOs can be generated by gigahertz magnetization precession[8] or megahertz vortex gyration[10,12] depending on the applied magnetic field.[13] In contrast to precession-based STOs, vortex oscillators exhibit multi-octave frequency tunability[14] and relatively high output power,[13] resulting in significant interest from applied and fundamental perspectives. While the output power of vortex auto-oscillations can be more than two orders of magnitude larger than precession-based oscillations in the same device,[13] the output power remains too low for technological applications.[6] At the same time, it is recognised that the linewidth of STOs can be undesirably large, owing to the frequency and phase noise of the auto-oscillations.[15,16,17,18,19,20] This has led to a number of studies of the mutual synchronization of multiple STOs[21,22,23,24,25,26,27,28,13,29,30] and phase-locking of STOs to an injected microwave (RF) current[20,31,32,33,34,35,36] in an attempt to enhance their output power and phase stability.

The synchronization and locking of STOs depends upon their non-linear nature.[37] In this work the non-linear character is utilised to achieve super-harmonic injection locking (SHIL) of a nano-contact (NC) spin-torque vortex oscillator (STVO) subject to an injected RF current with a frequency $f_{RF}$ that corresponds to higher harmonics ($n \times f_0$) of the fundamental STVO auto-oscillation frequency $f_0$. SHIL is observed up to the fourth harmonic ($4f_0$) in microwave magneto-electronic measurements, while time-resolved scanning Kerr microscopy (TRSKM) is used to observe differences in the spatial character of magnetization dynamics generated by the STVO when injection-locked to either its fundamental frequency $f_0$, or the second harmonic frequency ($2f_0$) through SHIL. Previously, SHIL has been used in complimentary metal-oxide-semiconductor (CMOS) technology to construct an injection locked frequency divider (ILFD) that produces a phase-locked response at a fraction of the applied microwave signal.[38,39] The multi-octave tunability of the STVO allows SHIL to be observed at multiple subharmonics (*e.g.* $f_{RF}/2$, and $f_{RF}/3$) for a particular frequency of the injected RF current. Hence, this work demonstrates that STVOs may also find a potential application as nanoscale ILFDs that utilize SHIL.

Recently, SHIL in a STVO was demonstrated in a single magnetic tunnel valve nano-pillar geometry consisting of a double vortex configuration of the free and reference layers.[20] In a nano-pillar STVO the magnetization dynamics are confined by the physical boundaries of the pillar. In contrast, here we demonstrate SHIL in a nano-contact (NC) geometry in which the applied current was confined by the NC before entering an extended spin valve film. The electrical contact pads to the device were designed specifically for optical access to the extended film in order to explore differences in the spatial character of the magnetization dynamics when the STVO was either SHIL at $2f_0$, or locked at $f_0$.

## II. EXPERIMENTAL SET-UP AND SAMPLE DETAILS

In this study each STVO was formed from a single metallic NC fabricated on top of a microscale spin valve mesa. The nominal diameter of the NCs was 250 nm. A spin valve multilayer film, consisting of Pd(8)/Cu(15)/Co(8)/Cu(7)/Ni$_{81}$Fe$_{19}$(4.5)/Cu(3)/Pd(3) (thickness in nanometres), was deposited onto a Si/SiO$_2$(thermally oxidised, 1 μm) substrate using magnetron sputtering in an argon plasma from a base pressure of $<2\times10^{-8}$ Torr. Photolithography and argon dry etching were then used to define a spin valve mesa with lateral size of 16×8 μm$^2$. The Co, Cu(7), and NiFe layers formed the reference (RL), spacer (SL), and free layers (FL) of the spin valve respectively, Figure 1(a). The bottom Cu(15) layer ensured that a significant proportion of the current passing through the NC flowed perpendicular to the magnetic layers to provide sufficient STT for device functionality, where a positive current represents electrons moving from the RL to the FL. To fabricate the NC, a SiO$_2$(30) layer was first deposited on top of the mesa using plasma enhanced chemical vapour deposition. The NC was then defined using electron-beam lithography, followed by reactive ion etching.

A coplanar waveguide (CPW), designed with a characteristic impedance of 50Ω, was used to make electrical contacts to the NC, Figure 1(b). The CPW was formed using a sputtered Cu(1200)/Au(20) bilayer and a lift off technique, and was specifically designed to only cover approximately half of the spin valve mesa allowing for optical access to the exposed FL (capped with SiO$_2$), Figure 1(c) (*c.f.* CPW design in Reference [13]). The NC was located at the centre of the mesa, which approximately places it on the symmetry axis and close to the edge of the center contact pad (see arrow in Figure 1(c)). The thick Cu contact pad provided an effective heat sink for device operation up to a current density of ~$2\times10^8$ A/cm$^2$, while the Au cap allowed for good electrical contact to the external electronics using either Au wire bonds or microwave probes. Electrical contact of the CPW ground planes with the Cu/Pd cap of the mesa was achieved using two large (~$4\times2$ μm$^2$) rectangular vias fabricated using photolithography and the same reactive ion etch used to fabricate the NC in the SiO$_2$.

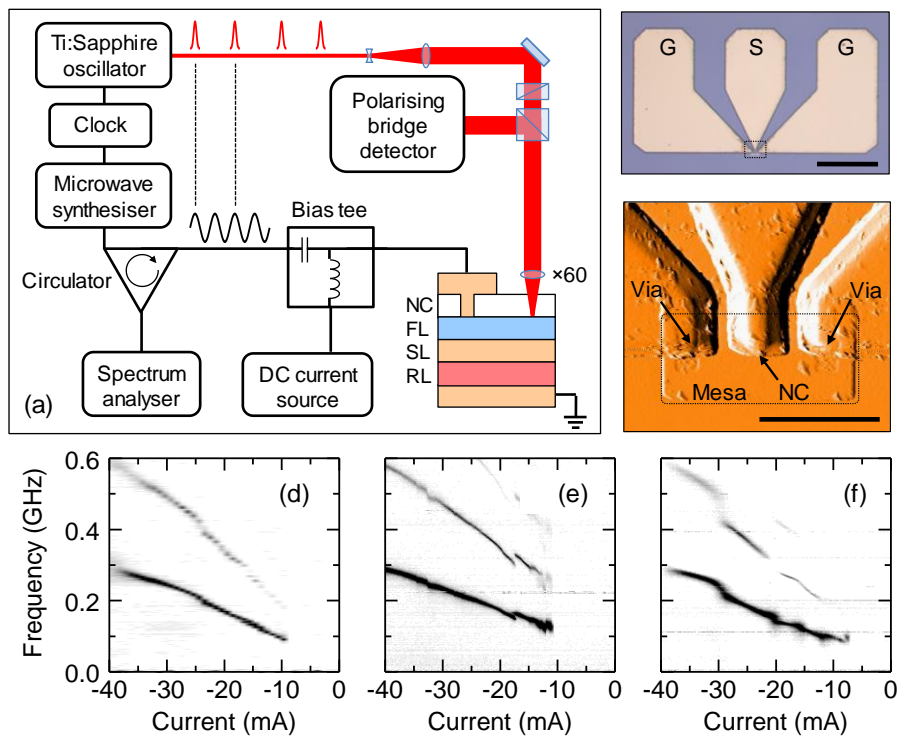

Figure 1. (Color online) (a) Experimental set-up of combined magneto-electronic measurements and TRSKM with coherent injection locking. (b) An optical microscope image of the CPW geometry of the ground (G) and signal (S) electrical contact pads (scale bar 100 μm). (c) A contact mode atomic force microscope image of the contact pad geometry on top of the spin valve mesa (dotted rectangle) that allows optical access to approximately half of the mesa (scale bar 10 μm). The location of the vias and the approximate location of the NC are indicated by arrows. In (d), (e), and (f) the STVO signal as a function of DC current only is shown for three single NC devices (NC1, NC2, and NC3 respectively) all with a nominal NC diameter of 250 nm. In (d-f) white and black correspond to 0 and 1 nV/Hz$^{1/2}$ respectively.

The experimental set-up for combined magneto-electronic measurements and TRSKM with coherent injection locking is shown in Figure 1(a). Measurements were performed on three single NC devices with nominally identical NC diameter (NC1, NC2, and NC3). The devices were fabricated simultaneously on the same wafer, yielding STVOs with similar free-running response, *i.e.* the response to a DC current only, Figure 1(d-f). Electrical measurements were performed by applying a DC current $I_{DC}$ to the NC via the DC port of a bias tee. The RF emission of the STVO was detected via the AC port using a spectrum analyser with a 10 dB pre-amplifier and a 100 kHz resolution bandwidth. A circulator was used to detect the RF emission of the STVO, while simultaneously applying an external RF current to the NC. The bandwidth of the circulator was 100 MHz (150 to 250 MHz), which was sufficient to observe the fundamental RF emission of the STVO.

To determine the optimum power $P_{RF}$ of the RF current for injection locking, preliminary electrical measurements were performed on device NC1.[40] Figure 2(a) shows frequency locking at 240 MHz around $I_{DC}$ = -30 mA for an injected RF current with $f_{RF}$ = 480 MHz and $P_{RF}$ = -10 dBm. As the DC current was swept, from immediately outside the locking range at $I_{DC}$ = -31.4 mA to the center of the locking range at $I_{DC}$ = -30 mA, an increase in amplitude (×5) and decrease in linewidth (×7) was observed, Figure 2(b). At the center of the locking range ($I_{DC}$ = -30 mA), a sweep of $P_{RF}$ from -30.5 dBm to -0.5 dBm, revealed that the maximum amplitude and minimum linewidth were observed at approximately $P_{RF}$ = -10 dBm, for which the locking range was almost a maximum.[41] When $P_{RF}$ < -30.5 dBm no frequency locking was observed, while at $P_{RF}$ = -0.5 dBm the STVO signal was no longer observed. To recover the STVO microwave signal, it was necessary to once again nucleate gyrotropic auto-oscillations of the vortex core in the absence of an injected RF current. Unless otherwise stated, $P_{RF}$ = -10 dBm for all injection locking experiments. For values of $f_{RF}$ that lay outside of the circulator bandwidth, the output power of the RF synthesiser was adjusted to compensate for transmission losses in the circulator to ensure that $P_{RF}$ ~ -10 dBm was injected through the NC for all values of $f_{RF}$.

For TRSKM measurements the microwave synthesiser was synchronised with the optical pulses of a femtosecond Ti:sapphire oscillator using the 10 MHz and 80 MHz outputs of a low phase noise master clock, respectively. Therefore, it was necessary for the RF frequency to be an integer multiple *m* of the laser repetition rate (*m*×80 MHz) in order to acquire stroboscopic scanning Kerr images. For example, the schematic of the laser pulse train and the RF waveform in Figure 1(a) corresponds to *m* = 2. For electrical measurements, it was not necessary to restrict the RF frequency to integer multiples of the laser repetition rate.

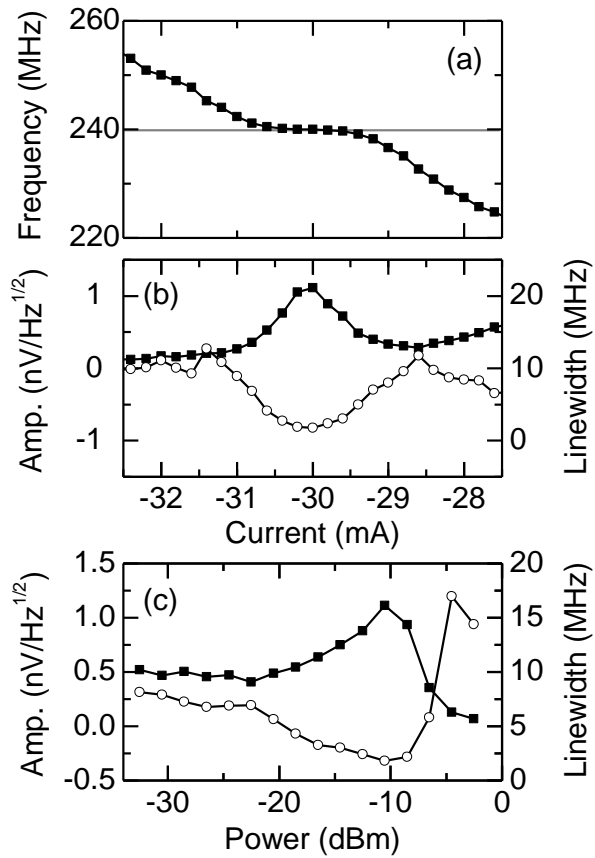

Figure 2. (a) The STVO frequency as a function of DC current of device NC1 shows frequency locking at $f_{RF}/2$ when $f_{RF}$ = 480 MHz. (b) The corresponding amplitude and linewidth reveal a maxima and minima close to the center of the locking range. (c) The amplitude and linewidth were found to be optimized when $P_{RF}$ was ~ -10 dBm.

TRSKM measurements were performed on device NC3 at a wavelength of 800 nm, at which the responsivity of the quadrant photodiodes of the polarizing bridge detector was approximately 90% of the optimal value of 6.5 A/W at ~920 nm. The probe beam was expanded ×5, linearly polarized, and focused to a diffraction limited spot onto the STVO mesa using a ×60 (0.85 numerical aperture) microscope objective lens. At a wavelength of 800 nm the spatial resolution was previously determined to be ~500 nm.[42] The quadrant photodiode, polarizing bridge detector was used to simultaneously detect two orthogonal in-plane components of the dynamic magnetization ($\Delta M_x$ and $\Delta M_y$) using the longitudinal magneto-optical Kerr effect, in addition to the out-of-plane component ($\Delta M_z$) using the polar Kerr effect. Time-resolved Kerr images were acquired at a fixed phase of the injected RF current by scanning the STVO mesa beneath the focused probe using a piezoelectric stage. Phase sensitive measurements were performed by phase modulating the RF current from 0 to 180° at

~3 kHz. Subsequently, vector-resolved Kerr signals, corresponding to $\Delta M_x$, $\Delta M_y$, and $\Delta M_z$, were recovered at the modulation frequency using three lock-in amplifiers.

**III. RESULTS AND DISCUSSION**

*A. Magneto-electronic measurements*

The free-running response of device NC2 to a DC current only is shown again in Figure 3(a). As the magnitude of $I_{DC}$ was increased from zero to 50 mA (not shown), the associated Oersted magnetic field curls the softer free layer magnetization around the NC, which favours a vortex within the otherwise uniform free layer magnetization. As the vortex is nucleated an anti-vortex is also generated and is understood to become pinned at defects within the free layer.[43] In zero applied magnetic field the onset of gyrotropic auto-oscillations of the vortex core was observed at $I_{DC}$ = -21 mA. As $I_{DC}$ was swept from -50 mA to -10 mA the fundamental frequency, or first harmonic ($f_0$), of gyrotropic auto-oscillations reduces from ~300 MHz to ~130 MHz, corresponding to more than one octave. A step transition in the frequency response can be seen around $I_{DC}$ = -15 mA. This is most likely due to a change in the core trajectory[44] as the equilibrium magnetization changes in response to the reducing amplitude of the Oersted field as the DC current approaches the 'cut-off' threshold at $I_{DC}$ = -10 mA. The second harmonic ($2f_0$) can also be seen, but appears to be weaker than might be expected due to the limited bandwidth of the circulator.

When an RF current with $f_{RF}$ = 400 MHz is injected, SHIL is observed, Figure 3(b). Two ranges of DC current around -25 mA, and around -15 mA, exhibit frequency locking at $f_{RF}/2$ and at $f_{RF}/3$, respectively. Surprisingly, the frequency locking at $I_{DC}$ = -15 mA ($f_{RF}/3$) appears to stabilise the transition observed in the free running response in Figure 3(a). In addition to multiple locking ranges, non-linear mixing (sum and difference signals) of the STVO fundamental response with the injected RF current leads to the presence of intermodulation modes.[26] The observed intermodulation modes in Figure 3(b) have frequencies of $2f_{RF} \pm f_0$ and therefore also show the signature of multiple frequency locking of the fundamental response.

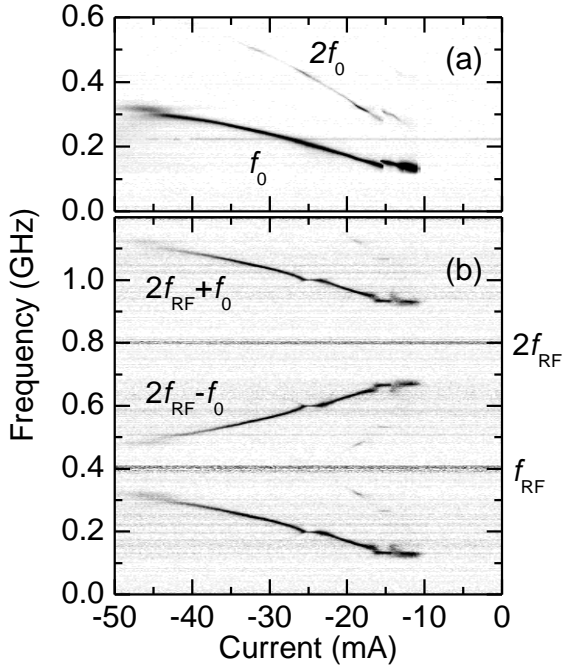

Figure 3. (a) The free-running response of device NC2 to a DC current only. (b) SHIL exhibited in device NC2 subject to an injected RF current with $f_{RF}$ = 400 MHz. In (a) and (b) the white to black greyscale represents 0 to 1 nV/Hz$^{1/2}$ and 0 to 0.5 nV/Hz$^{1/2}$ respectively.

Analysis of the RF emission reveals the expected increase in linewidth quality and enhanced amplitude in the locking range corresponding to $f_{RF}/2$, Figure 4(b) and 4(c) respectively. The response is similar to that observed in device NC1 (Figure 2(b)), which demonstrates device-to-device consistency of the injection locking behaviour. Figure 4(a) shows a direct comparison of the STVO fundamental frequency with and without an injected RF current with $f_{RF}$ = 400 MHz. As observed in Figure 3(b), two frequency locking ranges are observed and shown in greater detail in the insets of Figure 4(a). As the STVO frequency approaches the half-harmonic ($f_{RF}/2$) of the injected RF current, frequency pulling[45] is observed (Figure 4(a) and top-right inset), with a corresponding increase in linewidth and decrease in amplitude (Figure 4(b) and 4(c) respectively). When the STVO is locked at $f_{RF}/2$, a reduction in the linewidth of approximately one order of magnitude (from 5.2 MHz to 570 kHz) is observed at $I_{DC}$ = -24.8 mA, while an increase in amplitude of approximately 60% from 4.3 nV/Hz$^{1/2}$ to 6.8 nV/Hz$^{1/2}$ is observed. A similar behaviour of the linewidth and amplitude as a function of applied DC current is observed when the $f_0$ approaches $f_{RF}/3$, albeit with enhanced linewidth to either side of the locking range. However, when $f_0$ is locked at $f_{RF}/3$, the linewidth is approximately equal to that of the free-running STVO, while the amplitude is less than a third of the free-running value. While SHIL appears to stabilise the frequency around $I_{DC}$ = -15 mA, where a change in the equilibrium magnetic state may take place, the similarity of the locked and free-running linewidths at $f_{RF}/3$ may suggest that there is significant disturbance to the vortex core trajectory (frequency via orbital radius, and phase via position on the orbit) due to the competition between the change in the magnetic state and the onset of SHIL.

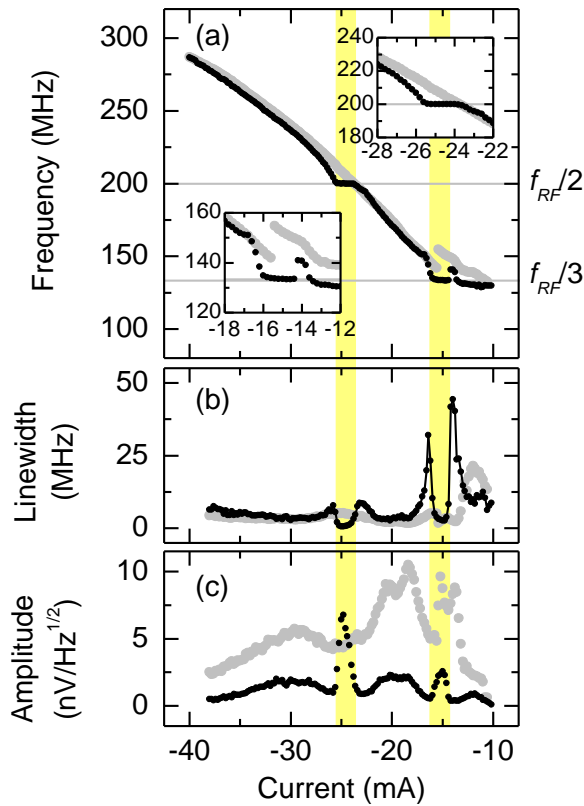

Figure 4. (Color online) (a) The free-running frequency response of device NC2 in response to the DC current only (grey filled symbols) and when exhibiting SHIL (black filled symbols) in response to an additional RF current with $f_{RF}$ = 400 MHz (black filled symbols). The locking range at $f_{RF}/2$ and $f_{RF}/3$ are shown in more detail in the top-right and bottom-left insets respectively. The linewidth (b) and amplitude (c) as a function of DC current for the free-running (grey filled symbols) and SHIL response (filled black symbols). In (a-c) the yellow bands correspond to the locking range.

The effect of the RF frequency on the efficiency of SHIL was explored. The RF frequency was varied from 100 MHz to 600 MHz in steps of 100 MHz. Sub-harmonic injection locking, where the RF frequency corresponds to a sub-harmonic of the STVO fundamental frequency ($f_{RF} = f_0/2$) has recently been reported in a nano-pillar tunnel valve with vortices in the free and reference layers.[20] In our NC-STVO such frequency locking to an injected RF current with a frequency of 100 MHz was not observed. However, when $f_{RF} = f_0/2$ non-linear mixing of the STVO fundamental response with the injected RF current was observed as a series of intermodulation modes with frequencies $2nf_{RF} \pm f_0$ ($n$ = 1, 2, 3, 4, 5).[46]

In general, when the frequency of the RF current was equal to, or greater than, the STVO fundamental frequency, Figure 5(a), injection locking was observed, Figure 5(b). More specifically, locking was observed when the frequency of the injected RF current crossed either the fundamental frequency, or the frequency of a higher harmonic, with the latter resulting in SHIL. It can then be understood that sub-harmonic injection locking is not expected in our device since there is no sub-harmonic RF emission at half of the STVO fundamental frequency ($f_0/2$).[38]

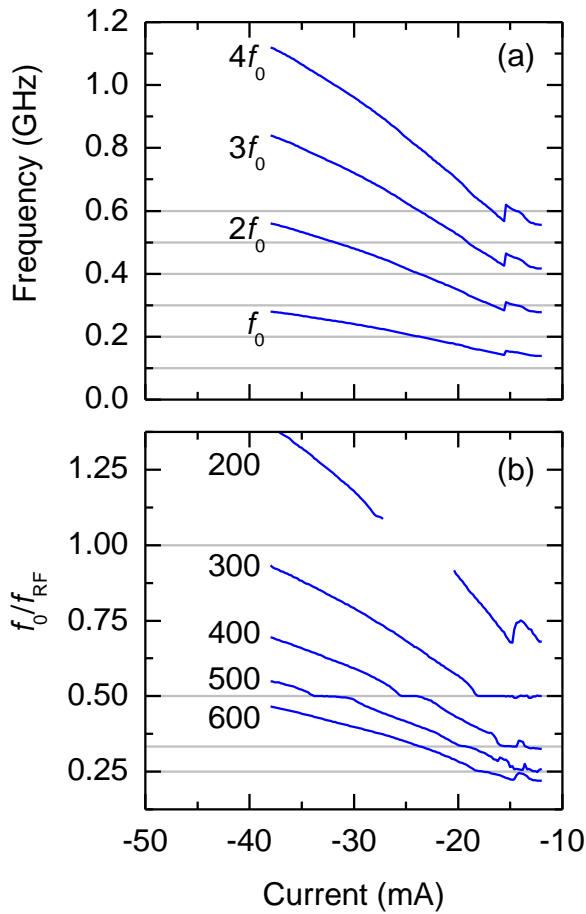

Figure 5. (Color online) (a) The free-running frequency response of device NC2 in response to the DC current only, and the calculated (i.e. $n \times f_0$) higher harmonics ($2f_0$, $3f_0$, $4f_0$), are shown (blue curves) overlaid with the frequencies of the injected RF current (grey curves). (b) SHIL of device NC2 by injecting an additional RF current with frequency in the range 200 MHz to 600 MHz (value/MHz listed).

When the fundamental frequency of the STVO was equal to that of the RF current ($f_0 = f_{RF}$), injection locking was observed around $I_{DC}$ = -23.6 mA that corresponds to the $f_0$ crossover with $f_{RF}$ at 200 MHz, Figure 5(a). Such frequency locking at the fundamental frequency has previously been reported by a number of groups.[20,32,33,34,35] Since the amplitude of the residual RF signal was at least two orders of magnitude larger than that of the STVO emission, the latter could not be reliably extracted around the locking range and have therefore been omitted from Figure 5(b). In this case only, the locking range was determined from the values of $I_{DC}$ at which the STVO response vanished beneath the RF signal.

When $f_{RF}$ = 300 MHz, locking to the fundamental response was not possible since the RF frequency was greater than the maximum fundamental frequency of the STVO, Figure 5(a). However, Figure 5(a) shows that the $f_{RF}$ coincides with that of the second harmonic ($2f_0$) at $I_{DC}$ = -16.6 mA, and again at $I_{DC}$ = -14.4 mA. SHIL then gives rise to frequency locking of the fundamental response around the crossover current with the second harmonic. At larger values of $f_{RF}$, SHIL was also observed. At $I_{DC}$ = -23.7 mA a crossover of $f_{RF}$ = 400 MHz with the second harmonic takes place, while at $I_{DC}$ = -15 mA a close approach of $f_{RF}$ to the third harmonic ($3f_0$) is sufficient for SHIL of the fundamental response to be observed. When $f_{RF}$ = 500 MHz two frequency locking events are again observed corresponding to a crossover with the second and third harmonics at $I_{DC}$ = -31.9 mA and $I_{DC}$ = -19.1 mA. Finally, when $f_{RF}$ = 600 MHz, frequency locking was only observed at $I_{DC}$ = -16.7 mA corresponding to a crossover of $f_{RF}$ with the fourth harmonic ($4f_0$), while locking at $f_{RF}/3$ (crossover with the third harmonic at $I_{DC}$ = -23.8 mA) was not observed. The observed locking ranges corresponding to SHIL are not centred on the crossover currents of the higher harmonics with the injected RF frequency, Figure 5(a) and 5(b). Frequency pulling tends to shift the locking range to a slightly larger magnitude of $I_{DC}$ than might be expected from the crossover values in Figure 5(a).

Figure 5(b) demonstrates that a STVO can be used as an ILFD, for which fractional locking is observed at $f_{RF}/2$ over a wide range of DC current. Furthermore, by tuning the DC current it is possible to access multiple fractions of the RF frequency at $f_{RF}/2$ and $f_{RF}/3$ in the same STVO for a single value of $f_{RF}$. A finer step in the value of $f_{RF}$ in device NC1 revealed evidence of fractional locking at $f_{RF}/2$, $f_{RF}/3$, and $f_{RF}/4$ between the onset and cut-off values of the DC current for $f_{RF}$ = 480 MHz.[40] This is only possible in STVOs for which the fundamental frequency of vortex gyration exhibits multi-octave tunability. In contrast STOs utilising higher frequency magnetization precession[13] would require bandwidths in excess of 40 GHz in order to exhibit SHIL at 20 GHz corresponding to $f_{RF}/2$, while the sub-octave tunability of such STOs would not permit frequency locking at $f_{RF}/3$ or $f_{RF}/4$.

Systems that exhibit SHIL are expected to show a reduced locking efficiency at higher harmonics.[20] To explore this further in our NC-STVO, the locking range, linewidth, and amplitude of each locking event observed in Figure 5(b) was extracted and plotted as a function of the locking fraction $f_0/f_{RF}$, Figure 6. The locking range was extracted from the derivative of the linewidth as a function of DC current for which the sharp reduction of the linewidth on approach to locking yields a peak in the derivative. The separation of the peaks was then taken as the locking range. One exception to this method was that used to extract the locking range when $f_{RF} = f_0$, for which the residual RF signal obscures the locked STVO emission, as described previously. In general there is an increase in locking range and amplitude as the locking fraction increases, which suggests that in general SHIL to the lower frequency harmonics is more efficient owing to their larger amplitude.[20] Similarly there is a general decrease in the linewidth as $f_0/f_{RF}$ increases (Figure 6(b)), again due to the enhanced efficiency of SHIL to the lower frequency harmonics. The efficiency of the non-linear process of SHIL depends upon the amplitude of the higher harmonics, and therefore decreases as the harmonic number $n$ increases.[20,47] In addition the linewidth of the higher harmonics of a free-running STVO has previously been shown to increase linearly as $n$ increases.[48] The increased linewidth of the higher harmonics may play a crucial role in the SHIL process whereby the dispersion of frequency and/or phase may allow the STVO to more readily lock to the injected RF current.[48] In contrast, a sharply tuned non-linear oscillator may not lock to an injected RF current, and so a minimum oscillator bandwidth may be required.[45]

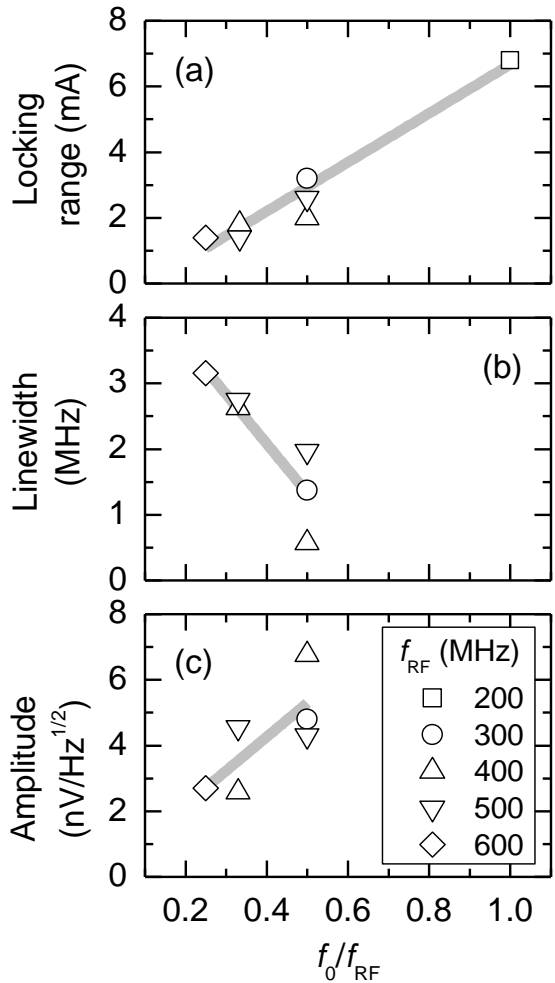

Figure 6. (a) Extracted locking range as a function of locking fraction. (b) Minimum linewidth within each locking event as a function of locking fraction. (c) Maximum amplitude within each locking event as a function of locking fraction. The grey lines only serve as guides to the eye.

Deviations from these general trends were observed. For example, the locking range was not observed to vary monotonically as a function of the RF frequency. When $f_{RF} = 500$ MHz and 300 MHz, the locking range at $f_{RF}/2$ was larger than that when $f_{RF} = 400$ MHz. The amplitude and linewidth of the free-running STVO emission may be responsible for these deviations. The linewidth is narrow at values of $I_{DC}$ corresponding to $f_{RF}/2$ locking events for $f_{RF}$ of 500 MHz and 300 MHz, Figure 4(b). At the same time, the amplitude of the free-running response was large in these regions, Figure 4(b). On the other hand, when $f_{RF} = 600$ MHz no SHIL was observed at $f_{RF}/3$ despite $f_{RF}$ crossing the third harmonic of the STVO at $I_{DC} = -23.8$ mA, Figure 5(a) and 5(b). This may be due to a change in the gyrotropic trajectory as implied by the onset of a sharp increase in the amplitude at a similar value of $I_{DC}$, Figure 4(b). As previously discussed the transitions in the core trajectory will take place as the Oersted field associated with the DC current reduces as the current is swept to the cut-off threshold. Corresponding changes in the symmetry of the trajectory as the DC current is reduced[44] may then lead to a variation in the efficiency of SHIL which then permits some locking events while others are more difficult to achieve.

*B. Time-resolved scanning Kerr microscopy measurements*

While injection locking can be used to enhance the output power and phase stability of individual STVOs, phase-locking of multiple STVOs can be achieved by injecting the same RF current, as demonstrated for a pair of tunnel valve nano-pillar STVOs.[33] In NC-STVOs it is anticipated that the magnetization dynamics within magnetic layers shared by multiple oscillators can be used to achieve mutual synchronisation resulting in enhanced output power and phase stability.[25] In addition to studying SHIL in electrical measurements, TRSKM was used to explore the differences in the magnetization dynamics generated by a single NC-STVO that is injection locked at its fundamental frequency $f_0$, and locked via the second harmonic ($2f_0$) using SHIL. To perform stroboscopic TRSKM measurements in these two locked states, it was necessary to set the frequency of the RF current to 160 MHz and 320 MHz respectively ($m = 2, 4$) as described earlier.

TRSKM images acquired from device NC3 are shown in Figure 7. The images represent the change in the dynamic magnetization due to the 180° phase modulation of the RF current while the STVO is phase-locked. In TRSKM measurements a small in-plane bias magnetic field of ~3 mT was applied to ensure that the FL returned to the same equilibrium state necessary for successful stroboscopic scanning Kerr imaging. The applied field led to an increase in the nucleation current to a value of $I_{DC}$ = -28.1 mA. When the fundamental frequency of the STVO was locked to an RF current with $f_{RF}$ = 160 MHz, localized dynamics were observed. The dynamics extend outside of the perimeter of the center contact pad in the vicinity of the NC, as indicated by location A in the reflectivity image of Figure 7(a). The observed dynamics have a complicated character.[49] The phase modulation of the RF current leads to the modulation of the vortex core position between two points on its trajectory that are separated by 180° At a particular phase of the RF current, the observed dynamics are interpreted as the deflection of the magnetization between these two dynamic states.[50] When the STVO is instead locked using SHIL at $f_{RF}/2$ = 160 MHz, where $f_{RF}$ = 320 MHz, the spatial character of the magnetization dynamics (Figure 7(b)) reveal significant differences to that observed in Figure 7(a). The strong contrast of the localized dynamics in Figure 7(a) is no longer observed, however weaker contrast with a different spatial character is observed at location A.[51] This can be interpreted as a different trajectory, *e.g.* orbital radius, of the vortex gyration when phase-locked at $f_0$ or $2f_0$. Movies of the magnetization dynamics at 320 MHz do not reveal any dynamics in the extended film that oscillate at 160 MHz. Therefore the locked STVO signal at $f_{RF}/2$ = 160 MHz is likely to correspond to vortex dynamics confined beneath the NC. Furthermore, the dynamic feature to the left of the center contact pad (indicated by white arrow) in Figure 7(a), and apparent absence in Figure 7(b) suggests that SHIL may also effect the equilibrium magnetization. The observed feature (arrow) in Figure 7(a) is understood to be an antivortex that has moved further from the NC, but has been trapped by stray electromagnetic fields generated by the DC current in the CPW structure. Micromagnetic simulations have shown that the anitvortex can move away from the NC and become pinned at the edge of the simulated mesh during the vortex-antivortex nucleation process.[43]

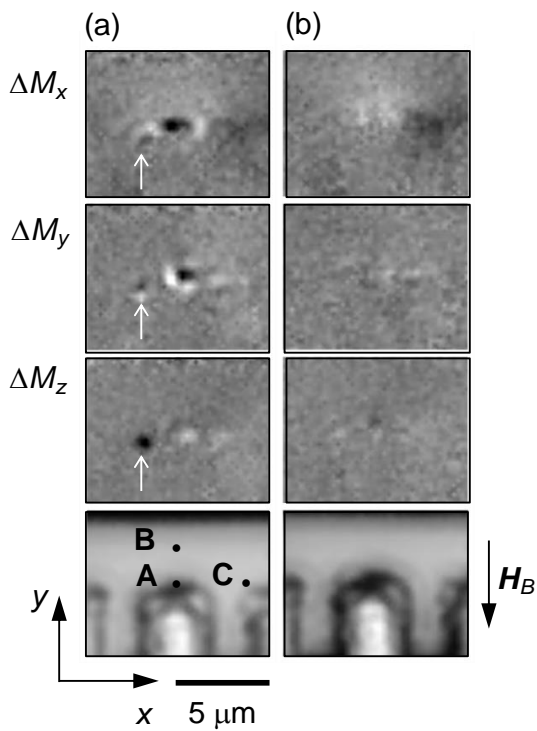

Figure 7. (a) TRSKM images corresponding to the change in three orthogonal components of the dynamic magnetization $\Delta M_x$, $\Delta M_y$, and $\Delta M_z$ for the case when the STVO is locked at $f_0 = f_{RF}$ (a) and at $f_0 = f_{RF}/2$ (b) using SHIL. The bottom image is a reflectivity image of the device. All four images in (a) and (b) were acquired simultaneously. In the Kerr images of $\Delta M_x$ and $\Delta M_y$, the greyscale represents the change in magnetization where black and white represent $+M_S/2$ and $-M_S/2$ respectively.

It should be noted that in Figure 7(a) and 7(b) the phase of the RF current is similar and can be determined by the contrast of the dynamics observed far from the NC, which is observed to be closer to white in the positive *y*-direction (location B), and closer to black in the positive *x*-direction (location C). These 'far-field' dynamics are the result of a small reorientation of the quasi-uniform magnetization far from the NC due to the Oersted field associated with the RF current passing through the mesa and the CPW structure. It should be noted that when the DC current is reduced to zero (not shown), only the far field dynamics are observed (not shown), confirming that the localized dynamics are related to the spin-torque excitation by the DC current.[49]

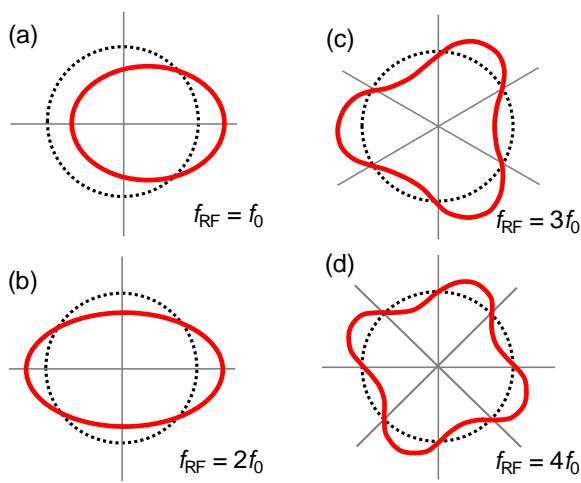

Figure 8. (Color online) Schematic illustration of the vortex core trajectory in the absence of an injected RFinjected RF current (dotted circle), and in the presence of an RF current (solid red trajectory) with frequency equal to the first (a), second (b), third (c) and fourth (d) harmonic of the free-running orbit.

While the trajectory of the vortex core in a NC-STVO may be complicated by the quasi-uniform equilibrium magnetization and the necessary presence of an anti-vortex within the FL, the trajectory can be considered in a simplified way for the case of SHIL. In a free-running NC-STVO, the vortex core can be assumed to occupy a concentric trajectory that is centred on the NC, while the Oersted field associated with the DC current acts as a concentric restoring potential that confines the core to its orbit.[34] When an RF current passes through the NC, an RF amplitude modulation of the current is introduced. While the modulation amplitude is typically an order of magnitude smaller than that of the DC current, the modulation of the Oersted field modulates the restoring potential. It can then be considered that the vortex core is periodically confined more tightly to the NC, or allowed to move further away. Figure 8 shows schematic illustrations of this interpretation. The frequency and phase noise associated with the vortex gyration allows its motion to phase-lock to an RF Oe-field that has frequency corresponding to the higher harmonics of the free-running response. This is shown for the first, second, third, and fourth harmonic in Figure 8(a), (b), (c) and (d) respectively, for which the trajectory of the free-running response is shown by the circular dotted line. The trajectory corresponding to $f_{RF} = f_0$, $2f_0$, $3f_0$, and $4f_0$, in Figure 8(a-d), show 1, 2, 3, and 4 periods of enhanced and diminished confinement (radius of orbit) within a single orbit at the fundamental frequency. One should note that this interpretation is an oversimplification of a realistic NC-STVO. It does not take into account the non-uniform equilibrium magnetization, complicated core trajectories in the presence of bias magnetic fields and from interlayer coupling, and asymmetries in the current distribution. This is quite clear from the direct observation of the dynamics using TRSKM, Figure 7. In the simple framework of Figure 8, one might expect to observe strong localized contrast in the vicinity of the NC in the Kerr images of Figure 7(b). However, the Kerr images suggest that the dynamics in the immediate vicinity of the NC are very different for the case when the STVO is locked at $f_0$ and $2f_0$.

**SUMMARY**

In summary, super-harmonic injection locking has been observed in a NC-STVO. The non-linear nature of the STVO allowed SHIL to be demonstrated up to (but not limited to) an injected RF frequency that was more than twice the maximum frequency of the free-running STVO. The SHIL locking events were shown to be located at values of the applied DC current that correspond to a crossover of the frequency of the injected RF current with that of the higher harmonics of the STVO. While there was a general correlation of the locking range, linewidth, and amplitude with the locking fraction, deviations from these trends were found to be related to variations in the linewidth and amplitude of the free-running response. TRSKM revealed that the spatial character of the magnetization dynamics in the vicinity of the NC was significantly different when the STVO was locked to an RF current with frequency corresponding to the first and second harmonics of the STVO. While the spatial character of the vortex core trajectory cannot be detected beneath the NC, a simplified interpretation of its gyration was presented. Specifically, a RF Oersted field, with frequency corresponding to one of the higher harmonics of the free-running STVO, can couple by perturbing the restoring potential experienced by the core. STVOs subject to SHIL may find future application as nanoscale injection locked frequency dividers. This work demonstrates that in a particular device, multiple locking fractions can be accessed for a particular RF frequency owing to the multi-octave tunability of the STVO.

**ACKNOWLEDGMENTS**

The authors gratefully acknowledge the financial support of the Engineering and Physical Sciences Research Council under grants EP/I038470/1 and EP/K008501/1, the Royal Society under grant UF080837, the Swedish Research Council (VR), the Swedish Foundation for Strategic Research (SSF), and the Knut and Alice Wallenberg Foundation (KAW).